\documentclass[manuscript]{aastex}

\slugcomment{Not to appear in Nonlearned J., 45.}

\shorttitle{SDSS\,J085338.27+033246.1: NLS1 and Post-starburst}
\shortauthors{Wang et al.}

\begin{document}


\title{Evolutionary Implications from SDSS\,J085338.27+033246.1: A Spectacular Narrow-Line
Seyfert 1 Galaxy with Young Post-starburst\
  }


\author{J. Wang\altaffilmark{1} and J. Y. Wei\altaffilmark{1}}
\affil{ National Astronomical Observatories, Chinese Academy of Science}

\email{wj@bao.ac.cn}



\begin{abstract}

We analyze the physical properties of post-starburst AGN SDSS\,J085338.27+033246.1 
according to its
optical spectrum and discuss its implications on AGN's evolution. 
The spectra PCA method is developed to extract emission lines and absorption 
features from the total light spectrum. 
The emission-line analysis indicates that the object can be classified as a 
NLS1 with FeII/H$\beta_{\rm{B}}=2.4\pm0.2$, large Eddington ratio ($\sim0.34$),
small black hole mass ($\sim1.1\times10^{7}M_{\odot}$) and intermediately strong 
radio emission. A simple SSP model indicates that the absorption features 
are reproduced by a $\sim$100 Myr old starburst with a mass of
$\sim7\times10^{9}M_{\odot}$ rather well. 
The current SFR $\sim3.0M_{\odot}\rm{yr^{-1}}$ inferred from the [OII] emission 
is much smaller than the past average SFR$\sim70M_{\odot}\rm{yr^{-1}}$, however.
The line ratio diagnosis using the BPT diagrams indicates that the narrow 
emission lines are almost entirely emitted from HII regions. 
We further discuss a possible evolutionary 
path that links AGN and starburst phenomena.

\end{abstract}




\section{Introduction}

In recent years, accumulating evidence supports the hypothesis that AGN reflects 
an important stage of galaxy formation. Observations reveals the fact that AGN activity 
and star formation frequently occur together (e.g., Cid Fernandes et al. 2001; 
Kauffmann et al. 2003; see reviews in Gonzalez Delgado 2002). 
An important clue of co-evolution of AGN and starburst is the fundamental, tight relationship 
between mass of supermassive black hole (SMBH) and the velocity dispersion 
of bulge where the SMBH resides in (e.g., Magorrian et al. 1998; Gebhardt et al. 2000; 
Tremaine et al. 2002, Greene \& Ho 2006). 
By analyzing the spectra of narrow-line AGNs from Sloan 
Digital Sky Survey (SDSS, York et al. 2000), 
Heckman et al. (2004) found that most of accretion-driven 
growth of SMBH takes place in the galaxy with relatively young stellar population. 
Recently, as essential relationships relating to AGN's activity, the Eigenvector 1 space 
(Boroson \& Green 1992, hereafter BG92) was found to be related with ages of 
the stellar populations as assessed by infrared color $\alpha$(60,25) (Wang et al. 2006).

It is then natural to ask a question ``What is a young AGN?''. Mathur (2000) argued that 
Narrow-line Seyfert 1 galaxy (NLS1) with high Eddington ratio and small black hole mass 
might be a young AGN. If so, the co-evolution of AGN and starburst implies that NLS1s are 
expected to be associated with relatively young stellar populations.

To understand the elusive AGN-starburst connection, we need to carefully examine the 
properties of AGN's host galaxy. However, optical spectroscopic study is possible for only a few  
type I AGNs because the strong nuclear emission usually masks the relatively faint stellar components.
By fitting deep off-nuclear optical spectra 
Canalizo \& Stockton (2001) and Nolan et al. (2001) determined the ages of circumnuclear stellar
populations for a few IR and optical-selected QSOs respectively.
In the happy case, 
Brotherton et al. (1999) discovered a spectacular ``post-starburst quasar'' UN\,J1025-0040 with 
a $\sim$400 Myr old stellar population. Wang et al. (2004) identified SDSS\,J022119.84+005628.4
as a NLS1 (RFe=1.4$\pm$1.0) whose spectrum displays a clear Balmer jump. However, the authors did 
not perform further stellar population analysis.

In this paper, we perform a detailed examination of the physical properties of 
post-starburst AGN SDSS\,J085338.27+033246.1 and discuss its implications on study of AGN's evolution. 
The optical spectrum is extracted from SDSS Data Release 4 (Adelman-McCarthy et al. 2006).  
Its appropriate spectrum allows us to separate it into a
young stellar population and a typical NLS1 spectrum. In addition, the emission line profiles 
allows us to determine the location on the line ratio diagnostic diagrams.  
We begin with a brief description of the data reduction in \S2. The results and implications are
presented in \S3.   
The cosmological parameters:
$h=0.7$, $\Omega_{\rm{M}}=0.3$, and $\Omega_{\rm{\Lambda}}=0.7$ are adopted for calculations.

\section{Spectral Analysis}

%% In a manner similar to \objectname authors can provide links to dataset
%% hosted at participating data centers via the \dataset{} command.  The
%% second curly bracket argument is printed in the text while the first
%% parentheses argument serves as the valid data set identifier.  Large
%% lists of data set are best provided in a table (see Table 3 for an example).
%% Valid data set identifiers should be obtained from the data center that
%% is currently hosting the data.
%%
%% Note that AASTeX interprets everything between the curly braces in the 
%% macro as regular text, so any special characters, e.g. "#" or "_," must be 
%% preceded by a backslash. Otherwise, you will get a LaTeX error when you 
%% compile your manuscript.  Special characters do not 
%% need to be escaped in the optional, square-bracket argument.

We extract the optical spectrum of SDSS\,J085338.27+033246.1 when we carry out 
a systematic search for broad line AGNs from the
objects listed in the catalogues \footnote{These catalogues can be downloaded from
http://www.mpa-garching.mpg.de/SDSS/.}
released by Kauffmann et al. (2003) and Heckman et al. (2004). 
Totally hundreds of such objects are extracted 
from the parent sample. The details of the search and properties of 
the sub-sample will be presented in subsequent papers. 
Among these objects SDSS\,J085338.27+033246.1 has unique
properties with strong enough typical AGN spectrum superposed on a spectrum of A-type 
star. It is coincident that we find the object has been identified as
a member of a sample of 74 post-starburst broad line AGNs by Zhou et al. (2005) when 
our paper is being prepared.

The observed spectrum is smoothed with a boxcar of 3 pixels ($\sim$5\AA) to enhance S/N 
ratio and precision of the spectra measurements. Standard IRAF 
procedures are adopted to reduce the raw data, including Galactic extinction correction 
(with $E(B-V)=0.048$ from NASA/IPAC Extragalactic Database (NED) and assuming $R_V=3.1$) 
and de-redshift to rest frame with $z=0.207881$.
The total light spectrum at rest frame is displayed in Figure 1 (the second line from top 
to bottom). The spectrum clearly shows a prominent Balmer jump at blue even diluted by an 
AGN continuum and high order Balmer series.

In the next step, the principal component analysis (PCA) method is developed to remove the 
stellar light from the observed spectrum (e.g., Li et al. 2005; Hao et al 2005). 
We build a library of stellar absorption spectra by applying PCA technique on 
standard SSP models developed by Bruzual \& Charlot (2003, hereafter BC03). The first 
seven eigenspectra are used to model the starlight component by their linear combination. 
In order to appropriate model the starlight component, our template contains the seven eigenspectra,
a power-law continuum, a FeII complex of AGN (adopted from BG92) and a Galactic extinction curve
(Cardelli et al. 1989). In order to avoid the distortion of low S/N, 
a $\chi^2$ minimizing is performed over the rest wavelength range 
from $\lambda$3700 to $\lambda$6800, except for the regions 
around the strong emission lines. The removal of the stellar component is illustrated in Figure 1 as well. 
Also plotted are the modeled FeII blends and emission-line spectrum of the AGN. The flux 
of the FeII blends measured between rest frame wavelength $\lambda$4434 and $\lambda$4684 is 
$(4.23\pm0.10)\times10^{-15}\ \rm{ergs\ s^{-1}\ cm^{-2} }$. The detailed 
comparison between the observed and modeled spectrum at blue is shown in the 
insert plot A. It is clear that 
the modeled spectrum reproduces the observed absorption features rather well.  
    
The AGN emission lines are modeled by SPECFIT task as described in Wang et al. (2005, 2006).
Briefly, each narrow line (e.g., H$\alpha$, H$\beta$, [NII]) is modeled by a Gaussian component. 
The intensity ratio of [NII] doublets is forced to equal to the theoretical prediction. 
The spectra modeling is schemed in the insert plot B and C for H$\beta$ and H$\alpha$
region, respectively. 
The upper limit of [OIII]$\lambda$5007 flux 
$F(\mathrm{[OIII]}) < 1.2\times10^{-16}\ \rm{ergs\ s^{-1}\ cm^{-2}}$ is estimated by integration 
rather than fit because the feature is very faint. In addition to these lines, the spectrum displays 
a broad emission feature at about 5880\AA. Based on its asymmetric profile, we identified this 
feature as HeI$\lambda$5876 contaminated by broad NaI\,D$\lambda\lambda$5890, 5896 emission. 
We find EW(HeI+NaI\,D)$\sim5.6\pm1.3$\AA\ and (HeI+NaI\,D)/H$\alpha_{\rm{B}}=0.11\pm0.02$. The NaI\,D emission 
was detected in a few AGNs (e.g., Veron-Cetty et al. 2004; Veron-Cetty et al. 2006 and references 
therein, Thompson 1991). 
The calculations indicated that the existence of large NaI\,D emission (NaI\,D/H$\alpha$=0.01-0.05)
is related to models with high density ($N_e\sim10^{11}\ \rm{cm^{-3}}$) and large column density 
($N_{\rm{H}}>10^{23.5}\ \rm{cm^{-2}}$) (Thompson 1991). The measured line properties are 
summarized in Table 1. The propagation of error is included in the uncertainties shown in parentheses.
All the uncertainties given in Columns (2) and (3) are caused by profile modeling.

%% In this section, we use  the \subsection command to set off
%% a subsection.  \footnote is used to insert a footnote to the text.


%% This section also includes several of the displayed math environments
%% mentioned in the Author Guide.

\section{Results and Discussion}

\subsection{Emission Line Analysis: A NLS1 with Intermediate Strong Radio Emission}

According to its spectra properties, we classify 
SDSS\,J085338.27+033246.1 as a NLS1 both because of the extremely faint [OIII]
emission ([OIII]/H$\beta<$0.05) and because of the relatively narrow H$\beta$ broad component 
(FWHM(H$\beta_{\rm{B}}$)=$1827.6\pm113.8\ \rm{km\ s^{-1}}$). Moreover, the measured 
RFe is as large as $2.4\pm0.2$, which is typical of NLS1s. 

We further calculate its black hole virial mass and Eddington ratio 
in terms of the H$\alpha$ broad component according to the equations in Greene \& Ho (2005):
\begin{equation}
M_{\mathrm{BH}}=2\times10^{6}\bigg(\frac{L_{\mathrm{H\alpha}}}{10^{42}\ \mathrm{ergs\ s^{-1}}}\bigg)^{0.55}
\bigg(\frac{\mathrm{FWHM(H\alpha)}}{1000\ \mathrm{km\ s^{-1}}}\bigg)^{2.06}M_{\odot}
\end{equation}
\begin{equation}
L_{5100}=2.4\times10^{43}\bigg(\frac{L_{\mathrm{H}\alpha}}{10^{42}\ \mathrm{erg\ s^{-1}}}\bigg)^{0.86}\ \rm{erg\ s^{-1}}
\end{equation}
 The intrinsic luminosity 
$L(\mathrm{H}\alpha_{\rm{B}})=(2.59\pm1.02)\times10^{42}\ \rm{ergs\ s^{-1}}$
is used for calculations because of the large intrinsic 
extinction $E(B-V)=0.42\pm0.18$. The extinction is calculated from the line ratio 
H$\alpha_{\rm{N}}$/H$\beta_{\rm{N}}=4.8\pm1.0$, assuming a Galactic extinction curve
and $R_V=3.1$ (Osterbrock 1989).
The inferred virial mass and Eddington ratio is $\sim1.1\times10^{7}M_{\odot}$ and $\sim0.34$, 
respectively. It is clear that all these properties (i.e., small black hole mass, large RFe value and 
large Eddington ratio) indicate the object should be potentially classified as a NLS1.

The radio radiation at 1.4GHz of the object was detected by NVSS (Condon et al. 1998)
and FIRST Survey (Becker et al. 1995). The map of the FIRST survey shows an unresolved 
source with a flux $\sim$4.04\,mJy and position 
$\alpha=\rm{08^h53^m38^s.269}$, $\delta=+03\symbol{23}32\symbol{19}46\symbol{125}.39$ (J2000). 
The discrepancy in position given by the optical and radio surveys 
is much less than 1\symbol{125} ($\sim$0\symbol{125}.3). We determined the 
extinction-corrected continuum flux that is contributed by AGN at rest 
wavelength $\lambda4400$ to be $0.16\pm0.02$mJy. 
If the radio emission is mainly due to AGN, the 
inferred radio loudness $R$ defined as $R=F_{\mathrm{radio}}/F_{\lambda4400}$ (Kellermann et al. 1989) is 
$\log R_{1.4\rm{GHz}}$=1.40 corresponding to $\log R$=1.12 for 5GHz radio flux\footnote{ 
$R_{1.4\rm{GHz}}=1.9R_{5\rm{GHz}}$ assuming a spectral shape $f_\nu\propto\nu^{-0.5}$ from optical to 
radio band.}.  Adopting the division $\log R=1$ between radio loud and quiet AGN, SDSS\,J085338.27+033246.1
is an object with intermediately strong radio emission.

We further calculate its radio luminosity 
$P_{1.4\rm{GHz}}\simeq(5.0\pm0.3)\times10^{23}\ \rm{W\ Hz^{-1}}$. 
This luminosity has same order of magnitude of 
the most radio-luminous starburst ($\log P_{4.85\rm{GHz}}\simeq22.3-23.4$, Smith et al. 1998).
It has long been known that the decimeter radio luminosity emitted from star-forming galaxy traces 
supernova rate of massive enough stars (i.e., $M\geq8M_{\odot}$). The lifetime of these massive 
supernova progenitors is $\approx10^{7.5}$yr. 
The radio supernova itself and its remnant has lifetime 
$\sim100$yr and $\sim2\times10^{4}$yr, respectively. The average SFR over the past 
$\sim10^{7.5-8}$yr therefore relates with the radio emission attributed to supernova as (Condon 1992)
\begin{equation}
\mathrm{SFR}(\geq5M_{\odot}) = \frac{L_{1.4\mathrm{GHz}}}{4.0\times10^{21}\ \rm{W\ Hz^{-1}}}\ M_{\odot}\rm{yr^{-1}}
\end{equation}
For the object, the estimated average SFR is $\sim70\ M_{\odot}\ \rm{yr^{-1}}$ (see Sect. 3.2 
for details). Using a Salpeter IMF (Salpeter 1955), this corresponds to a 
$\mathrm{SFR}(>5M_{\odot})\sim14M_{\odot}\ \rm{yr^{-1}}$. 
The inferred luminosity is therefore $\sim5.6\times10^{22}\ \rm{W\ Hz^{-1}}$ and 
is about one order lower than the total radio luminosity.

\subsection{Post-starburst: ``A-type star'' spectrum and residual star formation}

The ``A-type star'' spectrum indicates the age of the stellar population 
is no more than a few$\times10^{8}$ yrs. To interpret the spectrum of the young stellar population, 
we extract a series of  
spectra of SSP models of BC03 with a solar metallicity and with a Chabrier initial 
mass function. The extracted spectra are matched with the observed spectrum 
\footnote{In this study the spectrum used for 
age diagnosis is corrected for intrinsic extinction derived from the PCA modeling.} 
one by one within the 
wavelength range 3500-5000\AA. The spectrum red of $\lambda$5000 is not used because this part is 
obviously contaminated by an old population. Figure 2 illustrates the best match between 
the observed spectrum and a 0.1Gyr old SSP model. The model predicted mass of the starburst
is $\sim7.0\times10^{9}M_{\odot}$. This simple SSP model shows a past average SFR 
$\sim70\ \rm{M_{\odot}yr^{-1}}$.

However, recent studies 
suggested that the AGN and starburst activity perhaps not co-evolve simultaneously.
Schmitt (2001) suggested that the starbursts may predominate over AGNs in their earlier 
phase. Yan et al. (2005) recently suggested that most of 
SDSS post-starburst galaxies hold somewhat AGN like activities. 
The post-starburst is the phase in which the galaxy already ceased star formation at 
some recent epoch. The derived $\sim0.1$Gyr post-starburst in SDSS\,J085338.27+033246.1 
is somewhat similar with the $\sim0.4$Gyr post-starburst that was obtained by Brotherton et al. 
(1999) in UN\,J1025-0040. 

[OII]$\lambda$3727 line is widely used as an empirical indicator of ongoing 
SFR ($\leq10^7$yr) for emission line galaxy surveys (e.g., Gallagher et al. 1989; Kennicutt 1992;
Hippelein et al. 2003). Its practicability led to several calibrations and a great deal discussions
(e.g., Kewley et al. 2004; Kennicutt 1998; Yan et al. 2005).
The extinction-corrected flux of [OII]$\lambda3727$ doublets is
$(1.9\pm0.7)\times10^{-15}\ \rm{ergs\ s^{-1}\ cm^{-2}}$ (with its observed flux shown in Table 1), 
which yields a ratio $\log$([OII]/[OIII])$>$0.55.
Adopting the ratio $\log$([OIII]/H$\beta_{\rm{N}}$)$<$-0.65, the object is consequently
located in the region that represents star formation (Kim et al. 2006 and references therein).
It means that the [OII] emission ought to be almost entirely attributed to star formation (see below).

We use the calibration 
\begin{equation}
\mathrm{SFR}=7.9\times\frac{L_{\mathrm{[OII],42}}}{16.73-1.75[\log(\mathrm{O/H})+12]}M_{\odot}\rm{yr^{-1}}
\end{equation}
(Kewley et al. 2004) to estimate the SFR, where $L_\mathrm{[OII],42}$ is the luminosity of [OII]
emission in units of $10^{42}\ \rm{ergs\ s^{-1}}$. 
The SFR is inferred to be round about 3.0 $M_{\odot}\ \rm{yr}^{-1}$ by adopting the
assumption used in Ho (2005), i.e., the metallicity is twice of solar corresponding 
to $\log (\rm{O/H})+12=9.2$. Here, we obtain
a residual but quenched ongoing SFR as obtained in the optically selected QSOs by Ho (2005).
The distance-independent birthrate parameter $b$ is defined as the ratio of the current 
SFR to the average past one (Kennicutt et al. 1994). As a particular case, 
the object has extremely small value of $b\sim0.04$, i.e., 
has present SFR significantly lower than the past.

\subsection{Location on the BPT diagram}

The traditional BPT diagram (Baldwin et al. 1981)
is a powerful tool for diagnosing the origin of emission 
of narrow lines for emission-line galaxies.
The location of the object in the BPT diagram in which the 
line ratio [OIII]/H$\beta$ is plotted against [NII]/H$\alpha$
is marked by solid square in Figure 3 (left panel). 
The solid and dashed line shows the demarcation line 
between AGN and starburst galaxy defined by Kewley et al. (2004) and Kauffmann et al. (2003), respectively. 
The figure shows clearly that the position of the object is marginally above the 
threshold used to define representative HII region. The similar plot is shown in Figure 3 (right 
panel) but for [OIII]/H$\beta$ v.s. ([SII]$\lambda$6719+6731)/H$\alpha$. In this case, the position 
of the object falls into the region that represents starburst and is far below the 
demarcation line. The combination of these two diagnosis suggests the observed narrow emission lines 
are mainly produced by star-formation. 
What is the distribution in the BPT diagram for post-starburst 
AGNs is a further interesting question.

\subsection{Implications}
Here it is interesting to give more discussions on the elusive AGN-starburst connection. 
Wang et al. (2006) recently indicated that the well-documented E1 sequence most likely 
represents an evolutionary track of AGN. The track further implies that AGN with high Eddington 
ratio evolves to one with low Eddington ratio.    
This result also naturally implies that NLS1s that occupy one 
extreme end of E1 should be associated with relatively young stellar populations. Apart from the 
evidence mentioned above, for the IR-selected QSOs, Canalizo \& Stockton (2001) found the 
ages of circumnuclear stellar populations range from current star formation to $\sim300$Myr old. 
Subsequent spectroscopy studies indicated that about 70\%-100\% IR-selected QSOs are extremely 
or intermediately strong FeII emitters (e.g., Lipari et al. 2003; Zheng et al. 2002). Moreover,
Zhou et al. (2005) found that more than half of the post-starburst type I AGN fulfill the 
criterion for NLS1. All these evidence appear to reveal that a young AGN is most likely 
a strong FeII emitter with relatively narrow H$\beta$ profile. 
In contrast, old stellar populations ($\sim 8-14$Gyr) are 
found to dominate the off-nuclear ($\simeq5"$) stellar population for 
optical-selected QSOs (Nolan et al. 2001 and references there in). 

These accumulating clues lead us to propose a possible evolutionary scenario that links
both AGN and starburst phenomena. The massive starburst might be dominant in the earlier
evolutionary phase. Then the starburst passively evolves to A-type stars when 
the plenty gas falls into the center of a galaxy under the gravitational attraction 
of SMBH. At one stage about $\sim\rm{a\ few}\times10^{8}$yr after the beginning of the starburst, 
the SMBH begin to accrete matter with small black hole mass and high Eddington ratio.
The feedback of the AGN suppresses circumnuclear star formation at the same time 
(e.g., Kim et al. 2006).
The black hole mass grow substantially in the formed bulge in this short phase 
(e.g. Mathur et al. 2001; Mathur \& Grupe 2005). The duration of the phase 
is commonly estimated as $e$-folding time scale 
$t=4.5\times10^{7}(\frac{\eta}{0.1})(\frac{L}{L_{\rm{Edd}}})$yr (Salpeter 1964), 
where $\eta$ is the radiative efficiency. The differential growth of AGN and starburst 
allows the NLS1s below the $M_{\rm{BH}}-\sigma_{*}$ relation for inactive galaxies 
to approach normal $M_{\rm{BH}}-\sigma_{*}$ relation
(Mathur \& Grupe 2005; Mathur et al. 2001; Wandel 2002; Bian \& Zhao 2004).

The AGN then shines with decreasing accretion rate and 
insignificantly increasing black hole mass 
because of the consumption of the gas. The scenario  
implies that the AGN and host co-evolve marginally around the common $M_{\rm{BH}}-\sigma_{*}$ 
relation. The luminous AGN phase persists for typical AGN lifetime 
which is usually believed to be $10^{7-8}$ yr with however large uncertainties 
(see reviews in Matini 2004). During this phase, the young stellar population continuously
ages and its importance gradually fades. The underlying old stellar population (a few Gyr)
finally becomes dominant in the emission of host of an old AGN. In fact, in addition to 
the old stellar populations, relatively old post-starbursts (0.1-2Gyr) are 
frequently detected in off-nuclear regions of a few powerful radio-loud AGNs 
(e.g. Tadhunter et al. 2005 and references therein). 
After the luminous phase, AGN then passively evolves 
to a stage with low luminosity, or low radiative efficiency.   

In summary, this scenario could be described as following path: starburst $\rightarrow$ 
NLS1+post-starburst $\rightarrow$ luminous AGN with decreased Eddington ratio+old stellar population
$\rightarrow$ less luminous AGN.  

We believe future studies of AGN+post-starburst 
composite objects such as SDSS\,J085338.27+033246.1 are important for developing 
a rational scenario to interpret the AGN-starburst connection.

\section{Conclusions}
The physical properties of post-starburst AGN SDSS\,J085338.27+033246.1 are derived by analyzing 
its optical spectrum. It allows us to obtain results as following.  
The object can be identified as a NLS1 (([OIII]/H$\beta<0.05$, 
$\rm{FWHM(H\beta)=1827.6\pm113.8\ km\ s^{-1}}$ and RFe=2.4$\pm$0.2) associated with 
a post-starburst stellar population as identified from the 
size of the Balmer jump. A simple SSP model indicates
the starburst with a mass $7\times10^{9}M_{\odot}$ possibly took place 
$\sim1.0\times10^{8}$ yr ago. The [OII] inferred SFR is as small as
$\sim$3.0$M_{\odot}\rm{yr^{-1}}$ which is much smaller than the past average
one $\sim70M_{\odot}\rm{yr^{-1}}$. Its locations in the BPT diagrams show the 
HII region significantly contributes to the emission of the narrow lines.  
We further discuss its implications on the elusive AGN-starburst 
connection.

\acknowledgments

We would like to thank the anonymous referee for very useful comments and 
important suggestions.
We thank D. W. Xu, C. N. Hao, J. S. Deng and C. Cao for valuable 
discussions.
We are grateful to Todd A. Boroson and Richard F. Green for providing 
us the FeII template. 
This work was supported by the National Science Foundation of China (grant 105030005 
and 10473013). This research has made use of the NASA/IPAC Extragalactic Database, which 
is operated by JPL, Caltech, under contact with the NASA. 
The SDSS archive data is created and distributed by the Alfred P. Sloan Foundation.

\clearpage

%% the file, uncomment the \plotone, \plottwo, and
%% If you need a layout that cannot be achieved with \plotone or
%% \plottwo, you can invoke the graphicx package directly with the
%% \includegraphics command or use \plotfiddle. For more information,
%% please see the tutorial on "Using Electronic Art with AASTeX" in the
%% documentation section at the AASTeX Web site,
%% http://www.journals.uchicago.edu/AAS/AASTeX.
%%
%% The examples below also include sample markup for submission of
%% supplemental electronic materials. As always, be sure to check
%% the instructions to authors for the journal you are submitting to
%% for specific submissions guidelines as they vary from

%% This example uses \plotone to include an EPS file scaled to
%% 80% of its natural size with \epsscale. Its caption
%% has been written to indicate that additional figure parts will be
%% available in the electronic journal.

\begin{figure}
\epsscale{1.0}
\plotone{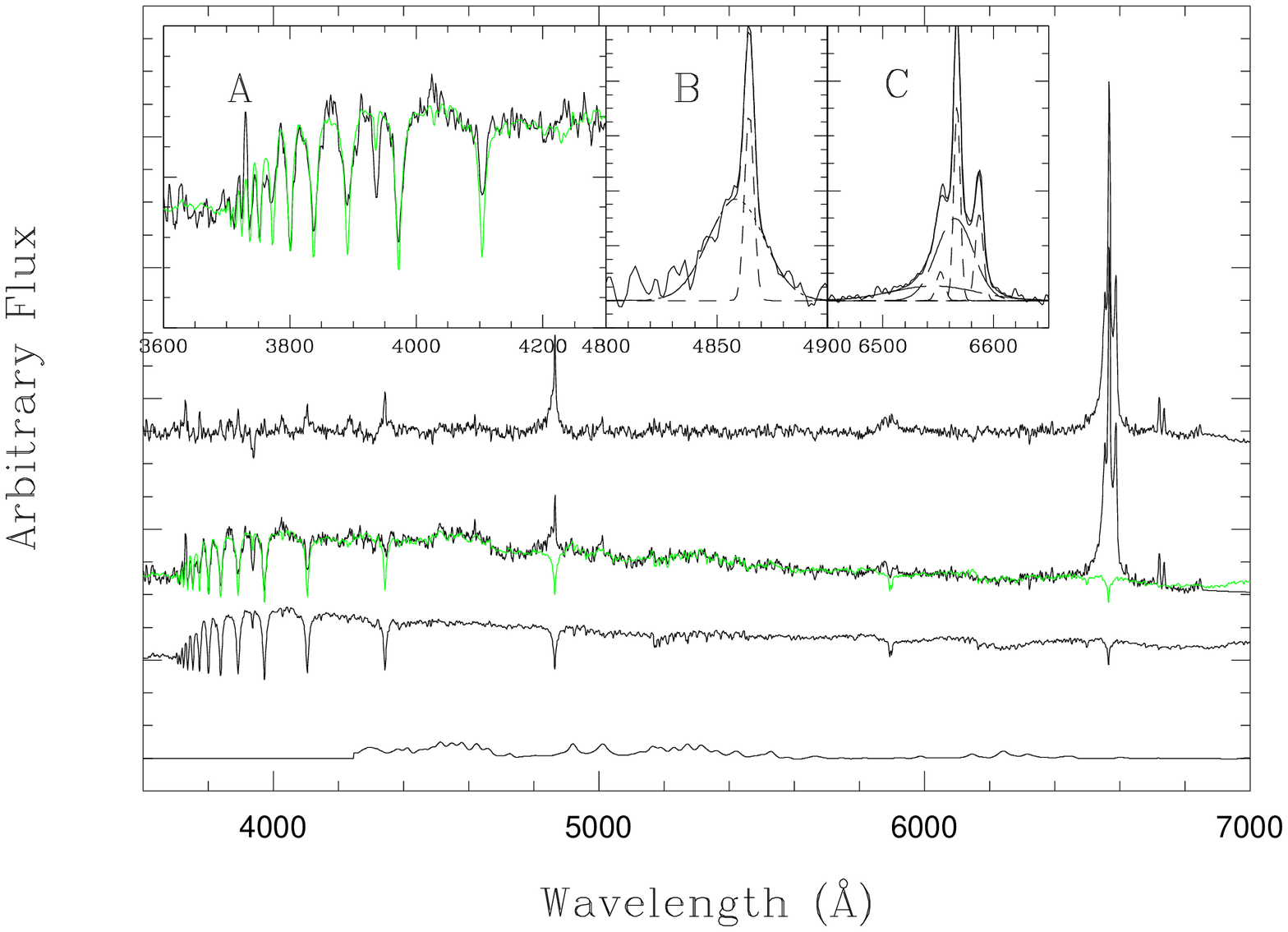}
\caption{Subtracting the stellar component using the linear combination of the 
7 eigenspectra and modeling of the emission line profiles. From top to bottom we 
plot the emission line spectrum, observed spectrum overlaid by the modeled spectrum, 
modeled spectrum of the host galaxy and modeled FeII blends. The insert plot A
shows detailed comparison between the observed and model spectrum at blue. 
The emission line fit is displayed in the insert plot B and C for H$\beta$ and 
H$\alpha$ region, respectively.}
\end{figure}

\clearpage

\begin{figure}
\epsscale{0.8}
\plotone{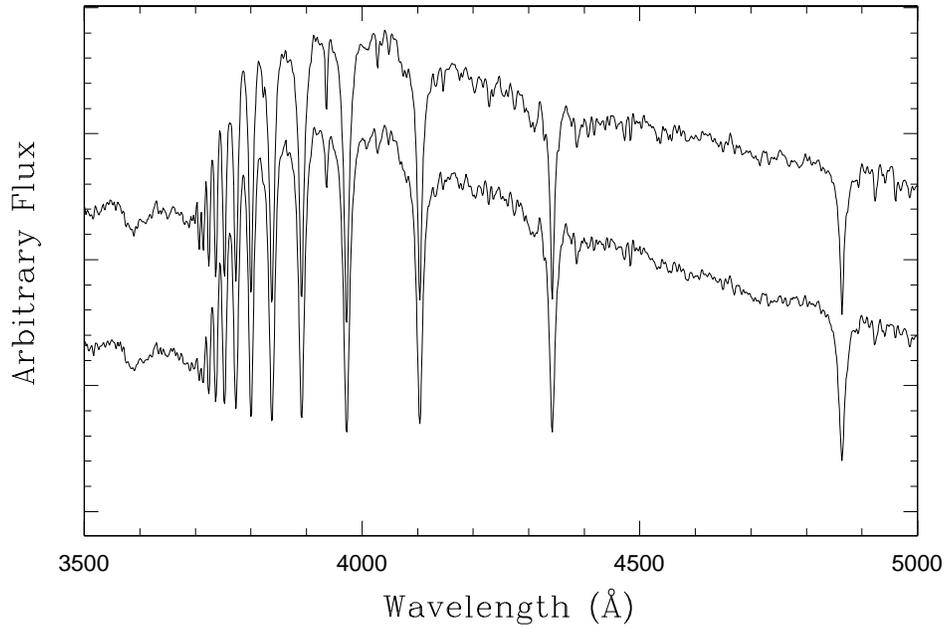}
\caption{An illustration of the match between the stellar spectrum 
and  a 0.1Gyr old SSP model with solar metallicity and Chabrier initial mass function. 
The SSP model is shifted downward arbitrary for visibility. 
}
\end{figure}

\begin{figure}
\epsscale{1.0}
\plotone{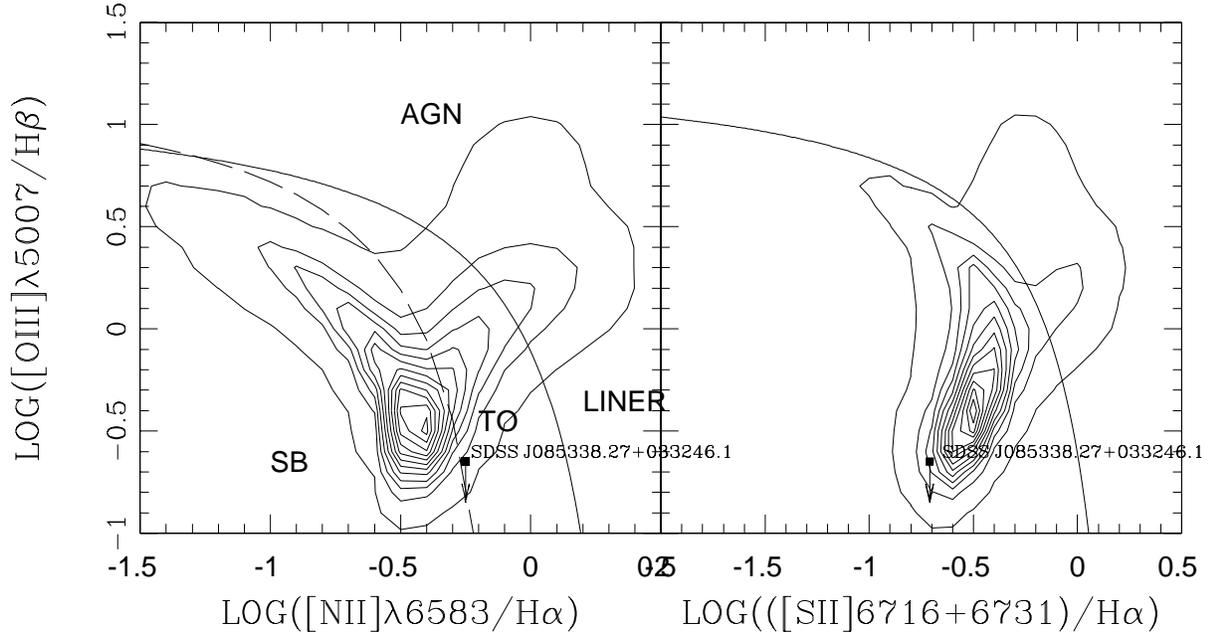}
\caption{\sl{Left panel}: \rm The location of SDSS\,J085338.27+033246.1 on the 
BPT diagram defined by line ratio [OIII]/H$\beta$
v.s. [NII]/H$\alpha$. The density contours are shown for typical distribution of 
the narrow-line galaxies described in Heckman et al. (2004) and Kauffmann et al. (2003). Only 
the galaxies with $S/N>20$ and the emission lines detected at at least 3$\sigma$ significance are plotted.  
The solid line is the demarcation line used to separate AGN from 
starburst by Kewley et al. (2004), and the dashed line by Kauffmann et al. (2003).
\sl{Right panel}: \rm The same BPT diagram but for [OIII]/H$\beta$ plotted against 
([SII]$\lambda$6716+6731)/H$\alpha$.}

\end{figure}

%% Here we use \plottwo to present two versions of the same figure,
%% one in black and white for print the other in RGB color
%% for online presentation. Note that the caption indicates
%% that a color version of the figure will be available online.
%%

%% This figure uses \includegraphics to scale and rotate the still frame

%% If you are not including electonic art with your submission, you may
%% mark up your captions using the \figcaption command. See the
%% No more than seven \figcaption commands are allowed per page,
%% so if you have more than seven captions, insert a \clearpage

%% Tables should be submitted one per page, so put a \clearpage before

%% Two options are available to the author for producing tables:  the
%% deluxetable environment provided by the AASTeX package or the LaTeX
%% table environment.  Use of deluxetable is preferred.
%%

%% Three table samples follow, two marked up in the deluxetable environment,
%% one marked up as a LaTeX table.

%% In this first example, note that the \tabletypesize{}
%% command has been used to reduce the font size of the table.
%% We also use the \rotate command to rotate the table to
%% landscape orientation since it is very wide even at the
%% Note also that the \label command needs to be placed

%% This table also includes a table comment indicating that the full
%% version will be available in machine-readable format in the electronic

\clearpage

%% Text for table notes should follow after the \enddata but before
%% the \end{deluxetable}. Make sure there is at least one \tablenotemark
%% in the table for each \tablenotetext.

%% If you use the table environment, please indicate horizontal rules using
%% Do not put multiple tabular environments within a single table.
%% The optional \label should appear inside the \caption command.

\clearpage

\begin{table}
\begin{center}
\caption{Properties of the emission lines of SDSS\,J085338.27+033246.1. Flux of 
each component is normalized to the flux of H$\beta$ narrow component 
$F(\mathrm{H}\beta_{\rm{N}})=(5.4\pm0.8)\times10^{-16} \rm{ergs\ s^{-1}\ cm^{-2}}$.
The uncertainties including propagation of errors are 
shown in the parentheses. 
}
\begin{tabular}{cccc}
\tableline\tableline
Line identification  & Flux ratio & FWHM \\
                     &            & ($\rm{km\ s^{-1}}$)\\
\tableline
H$\beta_{\rm{N}}$\dotfill & 1.0\dotfill & $301.8\pm38.0$\\
H$\beta_{\rm{B}}$\dotfill & $3.2\pm0.2(\pm0.6)$\dotfill & $1827.6\pm113.8$\\
$[\rm{OIII}]\lambda$5007\dotfill & $<0.22\dotfill$ & \dotfill\\
H$\alpha_{\rm{N}}$\dotfill & $4.8\pm0.4(\pm1.4)$\dotfill & $314.9\pm23.3$\\
H$\alpha_{\rm{B}}$\dotfill & $16.1\pm1.1(\pm3.1)$\dotfill & $1778.6\pm150.0$ \\
$[\rm{NII}]\lambda6583$ \dotfill & $2.7\pm0.3(\pm0.6)$\dotfill & $402.2\pm36.8$\\
FeII 4570\dotfill         & $7.8\pm0.2(\pm1.4)$\dotfill & \dotfill\\
$[\rm{OII}]\lambda$3727\dotfill & $0.56\pm0.02(\pm0.10)$ & \dotfill\\
$[\rm{SII}]\lambda$6716\dotfill & $0.59\pm0.06(\pm0.12)$ & $240.2\pm10.0$\\
$[\rm{SII}]\lambda$6731\dotfill & $0.35\pm0.01(\pm0.06)$ & \dotfill\\
HeI+NaI\,D             \dotfill & $1.85\pm1.24(\pm0.41)$ & \dotfill\\

\tableline
\end{tabular}
%% Any table notes must follow the \end{tabular} command.
\end{center}
\end{table}

%% If the table is more than one page long, the width of the table can vary
%% from page to page when the default \tablewidth is used, as below.  The
%% individual table widths for each page will be written to the log file; a
%% maximum tablewidth for the table can be computed from these values.
%% The \tablewidth argument can then be reset and the file reprocessed, so
%% that the table is of uniform width throughout. Try getting the widths
%% from the log file and changing the \tablewidth parameter to see how
%% adjusting this value affects table formatting.

%% The \dataset{} macro has also been applied to a few of the objects to
%% show how many observations can be tagged in a table.

%% You can append references to a table using the \tablerefs command.
%% Tables may also be prepared as separate files. See the accompanying
%% sample file table.tex for an example of an external table file.
%% To include an external file in your main document, use the \input
%% command. Uncomment the line below to include table.tex in this
%% sample file. (Note that you will need to comment out the \documentclass,
%% \begin{document}, and \end{document} commands from table.tex if you want
%% to include it in this document.)


%% The following command ends your manuscript. LaTeX will ignore any text

\end{document}